# Sustaining ALMA Science Through 2030
# A North American Perspective

Al Wootten
North American ALMA Science Center
NRAO
Charlottesville, Virginia 22903 USA
awootten@nrao.edu

*Abstract*— ALMA will sustain its transformational science through 2030 via an aggressive series of upgrades, for which an overview is provided.

I. INTRODUCTION

The Atacama Large Millimeter/submillimeter Array (ALMA)[1] provides astronomers with transformational spectroscopic sensitivity and imaging accuracy. ALMA's immense collecting area of over 6600m$^2$ distributed among 66 high precision telescopes are deployable over an extent up to 16 km on the lofty Chajnantor plain at 5000m elevation in the Atacama Desert of northern Chile. High resolution, combined with its current commodious spectral grasp of 8 GHz of spectrum in dual polarizations within its current 84-950 GHz range underlies its capabilities [1]. ALMA, an effort of 22 countries, has recently completed its fifth year of operation. Now the largest element in an even larger array with intercontinental baselines, an observing session expected to yield microarcsecond images was recently completed to investigate nearby Black Holes on the scales of their Event Horizons. ALMA has delivered over 1,000 datasets that have resulted in over 800 refereed publications. ALMA's scientific output has transformed millimeter astronomy. The ALMA Development Program sustains the pace of ALMA science through community-led studies and implementation of improvements to ALMA hardware, software, and techniques. During the upcoming decade through 2030, new capabilities will expand ALMA's envelope of exploration even further. ALMA will complete its 35-950 GHz spectral grasp and increase its spectral coverage and sensitivity within that commodious spectral window. Here we discuss hardware alone.

---

[1] ALMA is a partnership of ESO (representing its member states), NSF (USA) and NINS (Japan), together with NRC (Canada), NSC and ASIAA (Taiwan), and KASI (Republic of Korea), in cooperation with the Republic of Chile. The Joint ALMA Observatory is operated by ESO, AUI/NRAO and NAOJ.

II. CURRENT AND NEW INITIATIVES

Building on the successes of the initial suite of capabilities, ALMA will enhance them in several key areas in the decade leading to 2030. The ALMA Science Advisory Committee (ASAC) examined community input for potential new technical initiatives to sustain ALMA science. ASAC recommended four pathways for future development [2], distilled from the ALMA2030 summary of studies [3]. These included improving the archive, increased bandwidth (for finer resolution over a broader band and increased sensitivity and spectral coverage), higher angular resolution (longer baselines), and increased wide field imaging speed. The scientific advantages of these improvements include the ability for enhanced imaging of planetary disks, galaxy assembly, and chemical analyses of star-forming regions.

A. ALMA Correlator Upgrade

A staged upgrade to the ALMA 64-antenna has been proposed that will enlarge the number of channels and increase resolution by 8x, while improving spectral sensitivity by employing higher accuracy correlation. Scientifically, this will enable broader spectral windows to be deployed at a given resolution instantaneously, offering scientists nearly an order of magnitude broader suite of lines with which to investigate protoplanetary disks and protostellar clouds. The sensitivity enhancement is equivalent to the addition of eight additional antennas. When complemented with an upgraded digitization and frequency distribution system and receivers, the correlation capacity will be doubled to 8 GHz per polarization and sideband. This would double the current continuum sensitivity of the array enabling deeper images in less time, better calibration and access to a much broader swath of spectrum for line surveys or deep redshift searches. Adiitionally, an improved special use correlator for the four Total Power antennas is under construction by East Asia at KASI in Korea.

B. ALMA Receivers

**ALMA Band 1** covers 35-50 GHz and is the lowest frequency on ALMA, with the largest beamsize. Science goals for this band include evolution of dust in protoplanetary disks; masers; molecular gas at high-redshift; grain growth and spinning dust; molecular lines; the Sunyaev-Zel'dovich effect; Zeeman effect and polarization. Band 1 construction is under way as an East Asia project led by ASIAA in collaboration with NAOJ, U of Chile, NRAO, and HIA. The Critical Design Review and Project Review were held Jan 19-20, 2016 at ASIAA in Taiwan, and the Band 1 LO Critical Design and Manufacture Review (CDMR) was held and passed on December 12th 2016. The Manufacturing Readiness Review is upcoming; first units have been manufactured and tested. They provide $T_{rx} < 28.0$ K for 35.5 – 48.5 GHz (80% bandwidth) and <30K for the full band, well within specifications. It is expected that Band 1 will be available by 2022.

**AMA Band 2** (nominally 67-90 GHz) offers new frequency coverage with singular importance for the study of cold and icy deuterium-enhanced regions and redshifted CO. The highest level science goal, detection of CO at z~3 from a galaxy like the Milky Way, the J=3-2 line at 86 GHz, falls within this band. A project to build a prototype was approved by the ALMA Board in its March 2014 meeting. A single cartridge was built. The budget, project plan, and science vs. cost have been evaluated; the Preliminary Design Review was held on 30-31 May 2017. An independent project to combine ALMA Bands 2+3 has yet to be reviewed; published measurements of the performance of the LNA to be used suggest implementing the wider bandwidth will compromise sensitivity over that used for Band 2 NA by about 16%, jeopardizing the highest level science goal.

The **ALMA Band 5** receiver (covering 163-211 GHz) has been installed on ALMA, offered for science use in Cycle 5, and observations are expected to begin in March 2018. It was developed by the Group for Advanced Receiver Development (GARD) at Onsala Space Observatory, Chalmers University of Technology, Sweden, and was tested at the APEX telescope in the SEPIA instrument. The first production receivers were built and delivered to ALMA in the first half of 2015 by a consortium consisting of the Netherlands Research School for Astronomy (NOVA) and GARD in partnership with the NRAO, which contributed the local oscillator to the project. The performance of the receiver is considerably better than 50 K over the band.

### C. Phased ALMA Upgrades

ALMA Phasing Project Phase 2 is intended to further enhance and expand the capabilities of the ALMA Phasing System through new software development and commissioning, furthering the goal of achieving the highest sensitivity possible on the longest possible baselines on earth. The Project will provide enhanced sensitivity through improved delay management, will enable spectral line VLBI and an extended frequency range for VLBI Band 1 through ALMA Band 7, introduction of a "passive" phasing mode to enable VLBI on weak sources such as pulsars (recently detected with ALMA) and ALMA 'total-power' VLBI capability (compatible with subarrays)

### D. Increased Collecting Area

Additional antennas would restore ALMA to its originally planned complement of 64 or more, benefitting all science programs by increasing sensitivity, decreasing integration times and improving imaging fidelity. As a result of construction to these initial plans, much of the infrastructure for a 64 element array exists. Broadening the bandwidth increases sensitivity, but additional collecting area increases sensitivity over the narrow bandwidths of most spectral line emission. This is a critical need for the characterization of targets from protoplanetary disks, in which planets are shaped, to extremely distant galaxies, where the first heavy elements are forged in the first stars.

### E. Longer Baselines

ALMA can currently image nearby (<200pc) terrestrial planet-forming zones on its longest baselines at its highest frequencies. Often, however, the disks at those frequencies are opaque and their interior structure hidden. Dust optical depth is less at lower frequencies, allowing better characterization of interior structure. Developing an imaging capability matching the resolution of that at the highest frequencies requires longer baselines. However, the number of antennas, their placement and achievable baseline lengths require study, driven by the science required at higher low-frequency resolution. There are a number of logistical issues that also require thought for baselines more extensive than a few dozen km.

### F. Focal Plane Arrays

Focal plane arrays increase the field of view and are important for objects too extended for imaging in a single ALMA beam. Currently, many ALMA fields require multiple pointings for achieving their science goals, whether within the Solar System, molecular clouds, nearby galaxies or cosmologically distant groups or clusters of galaxies. While focal plane area is limited on ALMA, initial studies have suggested modest-sized arrays could be accommodated [4]. The science demands need to be studied to develop requirements on the preferred frequencies, the number of pixels and the necessary bandwidth. Arrays might be implemented on a subset of antennas, for instance the four in the Total Power array, for initial investigation, as interferometry with Focal Plane Arrays is a challenging instrumental goal.